# Two-Level Atom in the Resonant Field – Critical Remarks

Brazovskaja N.V, Brazovsky V.Ye.    braz@agtu.secna.ru
General Physics Dept, Altai State Technical University, Barnaul, Russia.

**Abstract**. We review the validity of the several representations of the two-level approximation.

## 1. Introduction

Quantum optics provides the ideal arena to deal with the interaction of radiation and matter. Indeed, by using standard techniques one easily gets a system of nonlinear coupled equations that govern, under the approximation of negligible damping, the interaction of a discrete set of field modes with an ensemble of atoms. The behaviour of a two-level particles in a resonant field - the area of investigation by a lot of scientists [1-66]. Nevertheless, as soon as one realizes the existence of fast and slow variables with a large difference in time scales, it is rather obvious that a simplified formulation can be derived from first principles, at least for some limiting cases. This is a very general description, but too involved to draw any immediate physical conclusion.

There are not full list of papers. We have chosen some of them for our analysis. We want to compare among themselves physical validity of initial rules of the various theories. The aim of this paper is to review, using the approach of quantum field theory, the consequences of the validity of the several representations of the two-level approximation. For this purpose we have chosen some works for the critical analysis of rules, contained in them, and statement of the own point of view on the appropriate problems.

The paper is so structured. In section II we describe the two-level approximation. In section III we analyze the Jaynes-Cummings model. In section IV we analyze the "superradiation Dicke". In section V we analyze the resonant dipole-dipole interaction.

## 2. Two-level atom

In a quantum electronics in a resonant case it is possible to take into account only two energy levels of a molecule or atom (hereinafter for a simplicity - atom) - at interaction with laser radiation [67]. From the methodical point of view the attractiveness of research of interaction of the system of two-level atoms with a resonance emission is encompassed by possibilities to construct visible, visual logical chain from the analysis of initial positions of the theory up to computed outcome of some effect observed experimentally.

The radiation-matter interaction currently used is based on some relevant approximations that are still well verified in the current experiments: Firstly, it is assumed that the dipole approximation holds, that is the wavelength of the radiation field being much larger than the atomic dimensions. Secondly, the rotating wave approximation (RWA) is always assumed, meaning by this that just near resonant terms are effective in describing the interaction between radiation and matter, these terms being also described as energy conserving. Thirdly, the intensity of laser irradiation. On the one hand it must be more then spontaneous emission, on the other hand it does not lead to energy-level splitting by the condition:

$$\frac{\mu E}{\hbar \Gamma} < 1 \,, \tag{1}$$

where $\mu$ is the matrix element of the dipole moment of the transition resonant with radiation, $E$ is the external field of laser irradiation, $\Gamma$ is the halfwidth of the transition.

With an external field $E(t)$, the Hamiltonian for complete system, atom plus electromagnetic field **E** in the limit of long wave-lengths and in the rotating-wave approximation, can be written as follows [2]:

$$H = H_f + H_a + V. \tag{2}$$

where $H_f$ is field energy, $H_a$ describe the energy of atom, perturbation is [67]:

$$V = -\mathbf{dE}, \tag{3}$$

This is a good scheme for the further quantum analysis of different real experimental situations. Then one must make any propositions for solving any task. But these propositions should be physically carefully justified.

### 3. Jaynes-Cummings model

The first proposition we will discuss is so called Jaynes-Cummings model [55]. The usually two-level system represent by a bipropellant wave function:

$$\Psi = \begin{pmatrix} \Psi_1 \\ \Psi_2 \end{pmatrix} \tag{4}$$

Then using this representation one write expression of perturbation (3) as follows:

$$V = -\mu\boldsymbol{\sigma}\mathbf{E}, \tag{5}$$

here $\boldsymbol{\sigma}$ expressed in terms of the Pauli matrices:

$$\sigma_x = \begin{pmatrix} 0 & 1 \\ 1 & 0 \end{pmatrix}, \quad \sigma_y = \begin{pmatrix} 0 & -i \\ i & 0 \end{pmatrix}, \quad \sigma_z = \begin{pmatrix} 1 & 0 \\ 0 & -1 \end{pmatrix} \tag{6}$$

Earlier we scored [68], that the representation of the two-level system by a bipropellant wave function and usage of the Pauli matrices for representation of electrical dipole moment (so-called Jaynes-Cummings model) is not enough justified. It mean that expression of perturbation (5) consist two vectors, and $\mathbf{E}$ is the polar vector, but $\boldsymbol{\sigma}$ is the axial vector. It multiplication lead to pseudoscalar, but energy must be scalar. Let one compare two interaction Hamiltonians: for magnetic field $g\boldsymbol{\sigma}\mathbf{B}$ (lead to so called Bloch equations, $\mathbf{B}$ is axial vector describing magnetic field) and for electric field $\mu\boldsymbol{\sigma}\mathbf{E}$. What is different mathematically? One can come to conclusion himself.

One can say that the introduction of Pauli matrices for the description of two-level atoms is only for convenience and has nothing to do with the transformation properties of a spin-operator. The Pauli-operators introduced can be considered as properly choosen dyadic products of the two states of the two-level system, it is not necessarily, to introduce explicitely the Pauli spin matrices.

All is right. A lot of physics can be derived by such an approximation and several recent experiments agree fairly well with a description given by the Jaynes-Cumming model describing a two-level atom interacting with a single radiation mode. On the other hand Jaynes-Cumming model is not widely justifies and is unconditionally suitable only for researches of effects of the first order. One can not use this model for the second order problems such as atom-atom interaction. One must have vector formulation of the interaction Hamiltonian for calculation the second order problems.

We proposed [68] other equation for describing two-level atoms in the resonant field:

$$(-\frac{\hbar}{i}\frac{\partial}{\partial t} - c(\alpha p) + \mu(\alpha E) - \beta^1\hbar\omega)\Psi = 0, \tag{7}$$

where used ordinary vector $\boldsymbol{\alpha}$, which can be written using Pauli matrices as follows:

$$\alpha = \begin{pmatrix} 0 & \sigma \\ \sigma & 0 \end{pmatrix}, \tag{8}$$

and $\beta^1$ as follows:

$$\beta^1 = \begin{pmatrix} 1 & 0 & 0 & 0 \\ 0 & 0 & 0 & 0 \\ 0 & 0 & -1 & 0 \\ 0 & 0 & 0 & 0 \end{pmatrix}. \tag{9}$$

Then the general solution of last equation can be written as follows:

$$\Psi = \begin{pmatrix} \Psi_1 \\ \Psi_2 \\ \Psi_3 \\ \Psi_4 \end{pmatrix} e^{+i\frac{pr}{\hbar}}. \tag{10}$$

### 4. Superradiation Dicke

There are a lot of papers contain interesting experimental and theoretical investigations which use expression "superradiation Dicke". We have no purpose to describe all this different situations. Let us consider initial proposition given by Dicke [1]: collective spontaneouse emission (superradiation Dicke) as correlation of spontaneouse dipoles interacting by own radiation field.

There are N identical two-level atoms in volume $V<\lambda^3$. Part of them there are exited. They give light together with intensity proportional $N^2$ through correlation spontaneous emitters by own radiation field. It must be mentioned, that it cannot include such phenomena as photon echo, as soon as photon echo take place through dipole correlation by external pulse radiation, not spontaneouse.

Let us consider the reality of this problem in usual representation. First position: every interaction must have such characteristic as energy, which is included in a nonrelativistic Hamiltonian. But usually using so-called Dicke Hamiltonian does not include any term of interatomic interaction energy. As soon as interatomic energy is absent, let us try to understand what does mean the expression "interaction throw own radiation field"? One else question is what the kind of theory are used? Is it nonlocal theory if N atom interact as one? But nonlocal theory is not exist now.

Quantum field theory treat an interaction of two particles as exchanging among it by particle of another sort. Elementary act of interaction of two atoms is follows: one atom emits photon and other atom absorb this photon. So, can two atoms interact on the interatomic distances $r$ more smaller then wavelenth $\lambda$ by exchanging resonant photon in the real medium?

Second position: every medium has such characteristic as amplification (or damping) coefficient $k$, given by Bouguer-Lambert-Beer low.

$$I = I_0 e^{-kr}, \tag{11}$$

here $I_0$ is initial intensity of radiation, $I$ is intensity of radiation after distance $r$ in medium. What does it mean from the photon point of view? There are any probability of stimulation emission (absorbtion), given by Einshtein coefficient B. As average value we can say that photon can be absorbed (interact with other atom) only throw distance $r=k^{-1}$ in the medium. For interaction two atoms by resonant spontaneouse emission one must have medium which parameter $k$ is more then $\lambda^{-1}$. For optical region one must have $k>100$ cm$^{-1}$. Is such medium real exist? But it is for interaction of two atoms. What about N atoms for Dicke model?

Third position. Kazancev and Surdutovich in 1969 [69] showed: if two atoms exchange by photon it lead to decorrelation of its irradiation, not to correlation!

As we can see such proposition as superradiation Dicke as correlation of spontaneouse dipoles interacting by own radiation field is mistake. Now any number effects of several physical nature named "effect Dicke". Good theoretical description of it – in future.

Last remark. In paper [70] one can read: "It was also shown by Dicke, that the system of two coupled two-level atoms can be treated as a single four-level system with modified decay rates. Note also that such model can be realized in a laboratory by two laser-cooled trapped ions, where the observation of superradiance and subradiance is possible." As just was showed the system of two coupled two-level atoms <u>can not be treated</u> as a single four-level system.

### 5. Dipole-Dipole interaction

Let us consider two atoms, driven by external monochromatic field. At first we will use classical point of view: dipole moment of the atom is given as follows:

$$d = \xi E \quad (12)$$

here $\xi$ is polarizability of the atom.

Dipole emits his own electric field $\mathbf{E_d}$, describing by expression:

$$E_d = \left\{ d\left(\frac{k^2}{r} + i\frac{k}{r^2} - \frac{1}{r^3}\right) - e_r(e_r d)\left(\frac{k^2}{r} + 3i\frac{k}{r^2} - \frac{3}{r^3}\right)\right\} e^{ikr} \quad (13)$$

here $\mathbf{e_r}$ is the unit vector in the direction $\mathbf{r}$.

Energy of dipole-dipole interaction is

$$U'(r) = -d_2 E_d \quad (14)$$

here $E_d$ is electric field of the first dipole in the point of the second dipole $d_2$.

Using such definition one can write for energy:

$$U'(r) = -\left\{ d_2 d\left(\frac{k^2}{r} + i\frac{k}{r^2} - \frac{1}{r^3}\right) - (e_r d_2)(e_r d)\left(\frac{k^2}{r} + 3i\frac{k}{r^2} - \frac{3}{r^3}\right)\right\} e^{ikr} . \quad (15)$$

Then one must take into account that both dipoles placed to the distance $r<\lambda$. This mean one must use Huygens-Fresnel principle: the electric field in this point is the superposition of all second emitters of the plane wave. In order of it one mast calculate average value of last expression by all possible directions of driven dipoles. The nature of the Huygens-Fresnel principle from the quantum point of view in our case is uncertainty principle.

Both dipoles are equal each other and have same directions. This mean we must calculate $<dd_2>$, not $<d><d_2>$, as in any papers. Then one can easily calculate average value as follows:

$$\langle U'(r) \rangle = -\frac{2d^2 k^2}{3r} . \quad (16)$$

Here we take into account $|d_2|=|d|$. At last we can take into account one more position: phase different of driven dipoles in the plane wave, which equal $\cos(\mathbf{kr})$. Lastly we can write for the energy of dipole-dipole interaction driven by external laser field as follows:

$$U(r) = -\frac{2d^2 k^2}{3r} \cos(\vec{k}\vec{r}) . \quad (17)$$

As we can see last expression does not consist terms $r^{-2}$ and $r^{-3}$. In the paper [3] dipole-dipole interaction was calculated using quantum theory, but authors forgot to calculate average value and obtained mistaken result.

Using representation (7)–(10) and quantum field theory we calculated dipole-dipole interaction driven by external laser field and obtained new low of interatomic interaction as follows [71 – 73]:

$$U = -\frac{\pi\mu^2 I_0 \beta}{12r(\Gamma_1^2 + \Delta_1^2)}(a\cos(kr) + b\sin(kr))\exp\left\{-kr\left|\frac{b - a\cdot tg(kr)}{a + b\cdot tg(kr)}\right|\right\}\cos(\vec{k}\vec{r}), \quad (18)$$

here frequency coefficients is $a = \frac{\Gamma_1 \Gamma_2^2}{\Gamma_2^2 + \Delta_2^2}$;

$$b = \frac{\Gamma_1 \Gamma_2 \Delta_2}{\Gamma_2^2 + \Delta_2^2} + (\Gamma_1 + \Gamma_2)\frac{(\Gamma_1 + \Gamma_2)\Delta_1 + \Gamma_1(\Delta_1 - \Delta_2)}{(\Gamma_1 + \Gamma_2)^2 + (\Delta_1 - \Delta_2)^2};$$

$\Delta_i = \omega_0 - \omega_i$; $\beta = \rho_2 - \rho_1$; $\rho_2$ and $\rho_1$ is population densities of upper and lower levels, $\Gamma_1$ and $\Gamma_2$ is dumping constants of first and second atoms (it may have different neighbour conditions), $\omega_0$ is central frequency of the laser field, $\omega_i$ is the frequency of the i-th atom (it may have different velosities), $I_0$ is intensity of laser irradiation.

We applied expression (18) to calculations of photocondensation, optical nonlinearity of two-level medium for $N_2$-laser, Nd-glass laser, light induced drift [71-74].